\documentstyle[12pt]{article}
\newcommand{\be}{\begin{equation}}
\newcommand{\bea}{\begin{eqnarray}}
\newcommand{\eea}{\end{eqnarray}}
\newcommand{\ba}{\begin{array}}
\newcommand{\ea}{\end{array}}
\newcommand{\ee}{\end{equation}}

\expandafter\ifx\csname mathbbm\endcsname\relax

\else

\fi
\def\flat{non-trivial flat connection }
\begin{document}
\begin{titlepage}
\hfill
\vbox{
    \halign{#\hfil   \cr
          IPM/P-2002/061 \cr  
          UCB-PTH-02/54  \cr
          LBNL-51767      \cr
          hep-th/0211285 \cr
          } % end of \halign
      }  % end of \vbox
\vspace*{8mm}

\begin{center}

{\Large {\bf Spacetime Quotients, Penrose Limits and } \\ }

\vspace*{2mm}

{\Large {\bf Restoration of Conformal Symmetry} \\ }

\vspace*{8mm}

{Mohsen Alishahiha$^a$,~ Mohammad M. Sheikh-Jabbari$^b$ and~Radu Tatar$^c$}
\vspace*{5mm}

{\it $^a\ $ Institute for Studies in Theoretical Physics and 
Mathematics (IPM)} \\
{\it P.O.Box 19395-5531, Tehran, Iran}\\
{\tt {e-mail: alishah@theory.ipm.ac.ir}}\\

\vspace*{2mm}

{\it $^b\ $ Department of Physics, Stanford University}\\
{\it 382 via Pueblo Mall, Stanford CA 94305-4060, USA}\\
{\tt {e-mail: jabbari@itp.stanford.edu}}\\

\vspace*{2mm}

{\it $^c\ $ Department of Physics, 366 Le Conte Hall}\\
{\it University of California, Berkeley, CA 94720}\\

\vspace*{2mm}

{\it and}\\

\vspace*{2mm}

{\it Lawrence Berkeley National Laboratory}\\
{\it Berkeley, CA 94720}\\
{\tt {email:rtatar@socrates.berkeley.edu}} 

\vspace*{1cm}

%%\maketitle
\end{center}                               

\begin{abstract}
In this paper we study the Penrose limit of $AdS_5$ orbifolds. The
orbifold can be either in the pure spatial directions or space and
time directions.  For the $AdS_5/\Gamma\times S^5$ spatial orbifold we 
observe that after the Penrose limit we obtain the same result as the 
Penrose limit of $AdS_5\times S^5/\Gamma$. We identify the corresponding 
BMN operators in terms of operators of the gauge theory on $R\times 
S^3/\Gamma$. The semi-classical
description of rotating strings in these backgrounds have also been 
studied. For the spatial $AdS$ orbifold we show that
in the quadratic order the obtained action for the 
fluctuations is the same as that in $S^5$ orbifold, however, the higher 
loop correction can distinguish between two cases.

\end{abstract}

\end{titlepage}

\section{Introduction}
In the past five years important results have been obtained relating 
closed 
and open string theories, in particular the AdS/CFT correspondence 
\cite{maldacena} had important impact into the understanding between these 
two sectors. The main problem in testing the conjecture
beyond the supergravity level is the fact that we do not know how to 
quantize string theory in the presence of RR-fluxes. 
More recently, the authors of \cite{bmn} have considered the sector of 
string states with large angular momentum along the central circle of 
$S^5$ (obtained in \cite{metsaev}) and have compared the quantum string 
oscillations with the spectrum of anomalous dimensions of field theory 
operators with large R-charge. This is the Penrose limit of the
$AdS_5 \times S^5$ which was also discussed in \cite{blau} 
yielding the maximal supersymmetric pp-wave.
The results of \cite{bmn} have been generalized to sphere orbifolds and 
other lower-supersymmetric backgrounds \cite{klebanov,mohsen,orb,kehagias} 
and the maps between string oscillators and definite operators of the 
field theory have been identified.

Another approach has been initiated in \cite{gubser} and expanded 
in \cite{FT} (see also \cite{russo,tse1}). It consists of choosing a 
classical solution of the string action and making an expansion in 
$R^2/\alpha'$ 
which are mapped into field theory energy spectrum and its 1-loop quantum 
corrections.

In the current project, we consider the issue of $AdS$ orbifolds for 
which the above map has not yet been identified. 
%(their Penrose limit was obtained in \cite{kehagias}). 
In this case, the field theories do not 
live on $R \times S^3$ but on $R \times S^3/\Gamma$ and the usual 
orbifold theory approach does not apply. The way to deal with these 
theories 
was outlined in \cite{horov} for field theories living on general $M/\Gamma$ 
manifolds where the orbifold action on $M$ is free and hence there is no 
singular locus. It is argued in \cite{horov} that the field theories 
should be modified by 
introducing a \flat 
%non-trivial flat connection 
which relates fields 
living on different patches of the manifold $M$. 

The main observation is that by taking the Penrose limit, the field 
corresponding to the boosted direction is required to live only on one 
patch and the covariant derivative is taken only with respect to the gauge 
field which corresponds to the same patch. 
More precisely, fields living in different patches lead to the twisted 
sectors for strings on plane wave orbifolds.
The fields which are orthogonal 
to the boosting direction are unchanged. In this limit, the theory maps 
naturally into the Penrose limit of the ${\cal N} = 2, S^5/Z_k$ orbifolds 
which were discussed in \cite{mohsen,orb}, and the different patches are 
mapped into the multiple coverings of $S^5/Z_k$. 

This map should also 
appear in the ``semi-classical'' approach of Ref. \cite{FT}. 
Performing the calculations we show that indeed the large 
 angular momentum sector of strings on $AdS_5\times S^5/\Gamma$ and 
$AdS_5/\Gamma\times S^5$ are the same
for the quadratic and quartic corrections. The conclusion is that the 
quadratic corrections are identical, but the quartic corrections
appear with a different sign. 

In section 2 we discuss the two types of orbifolds of $AdS_5$, one 
corresponding to a space orbifold and the other to a space-time orbifold 
(the complete discussion has appeared in \cite{Sunil}). We will 
concentrate 
in section 3 on the field theory dual to a space $AdS_5$ orbifold and 
leave the space-time orbifold for a future work. In section 4 we discuss 
the Penrose limits of the two types of $AdS_5$ orbifolds and the BMN 
operators for our case. In section 5 we apply the method of 
\cite{gubser,FT} to our cases and compare the quantum  corrections.

\section{Half supersymmetric orbifolds of $AdS_5$}
In this section we discuss quotients of $AdS$ spaces (see 
\cite{Sunil,gao,bl} for previous discussions).
The $AdS_5\times S^5$ space can be embedded in a twelve dimensional 
space with two time coordinates, i.e. the $AdS_5$ piece is sitting in a 
(4+2)-dimensional space and $S^5$ in a six dimensional Euclidean space. 

Let us first consider the $AdS_5$ part and let $z,u$ and $v$ be three
complex coordinate satisfying
\be
-|z|^2+|u|^2+|v|^2=-R^2\ .
\ee
In general the $AdS_5$ space has a $SO(4,2)$ global symmetry and one can 
orbifold the space by a (discrete) subgroup $\Gamma$ of that. Then as 
discussed in 
\cite{Sunil} {half} supersymmetric orbifolds of $AdS_5$ can be obtained 
in two different ways
\begin{enumerate}

\item Space orbifold
\[	
u\equiv \gamma u\ ,\ \ \  v\equiv \gamma^{-1} v\;,
\]
\item Space-time orbifold
\[
z\equiv \gamma z\ ,\ \ \  v\equiv \gamma^{-1} v\;,
\]

\end{enumerate}
where $\gamma\in \Gamma$ and for the case of ${\bf 
Z_k}$ orbifold $\gamma=e^{2\pi i/k}$. 

The case one corresponds to modding  out by a subgroup of $SO(4)\in 
SO(4,2)$ and hence the symmetry of the quotient space is $SU(2)\times 
U(1)$. This orbifold has a fixed line (the global time). The second one 
is obtained by quotienting with  $SO(2,2)\in SO(4,2)$ and has a
fixed circle and the final surviving symmetry is $SU(1,1)\times U(1)$.  

The metric for $AdS/\Gamma$ orbifolds is of course similar to the
simple $AdS_5$ metric, where some of the angular variables have been 
limited to range from zero to $2\pi/k$. However, for our purpose (taking 
the Penrose limit) it would be more convenient to choose the global 
coordinate system. Let us first focus on the first case, for 
which it is more convenient to use the following parameterization for
the complex coordinates
\bea 
z&=&R\cosh \rho\; e^{it} \cr
u&=&R\sinh \rho \;\cos\delta \;e^{i{\phi\over k}} \cr
v&=&R\sinh \rho\; \sin\delta \;e^{i({\phi\over k}+\gamma)}\ , 
\eea
where all the angular coordinates are now ranging from zero to $2\pi$ and
$\rho\in[0,\infty)$. The metric in the above coordinates takes the form
\bea\label{caseI}
ds^2&=&-dz{\bar dz}+du{\bar du}+du{\bar du}+ds^2_{S^5}\cr
&=&\alpha' R^2\bigg{[}-\cosh^2\rho\; 
dt^2+d\rho^2+\sinh^2\rho\;\bigg{(}d\delta^2+
{1\over k^2}\cos^2\delta \;d\phi^2\cr
&+& \sin^2\delta \;({1\over k}d\phi+d\gamma)^2\bigg{)}
+ \cos^2\alpha \;d\theta^2+d\alpha^2+\sin^2\alpha \;d\Omega^2_3\bigg{]}\ .
\eea
Here we have used a notation in which $R$ is dimensionless and $\rho$ is 
a radial coordinate 
transverse to time. The boundary of the spacetime, which is  at
$\rho =\infty$, is $R\times S^3/\Gamma$ 
%If we how rescaled the metric by $exp(-2\rho)$ and take the limit 
%$\rho \rightarrow \infty$,
%we obtain that the boundary of the above $AdS$ orbifold is $R\times S^3/Z_k$ 
and this is 
where the dual gauge theory resides. We note that although the bulk
$AdS_5/\Gamma$ has a fixed circle (which can be extended to a real line, 
the global time), for the boundary $\Gamma$ is freely 
acting on $S^3$ and hence there is no singular locus there. 
Due to orbifolding, the dual gauge theory is not Lorentz invariant. 
We would also like to recall that the $AdS$ orbifold preserves 16 
supercharges, and hence the dual gauge theory is a ${\cal 
N}=2,\ D=4$ theory.

For orbifolds of the second type, however, it is better to use the
parameterization:
\bea 
z&=&R\cosh \rho\; e^{i{t\over k}} \cr
u&=&R\sinh \rho\; \cos\delta \;e^{i({t\over k}+\phi)} \cr
v&=&R\sinh \rho\; \sin\delta \; e^{i\gamma}\ , 
\eea
and hence the metric reads as
\bea\label{caseII}
ds^2&=&\alpha' R^2\bigg{[}-{1\over k^2}\cosh^2\rho\;dt^2+
d\rho^2+\sinh^2\rho\;\bigg{(}d\delta^2+\sin^2\delta \;d\gamma^2\cr
&+& \cos^2\delta \;({1\over k}dt+d\phi)^2\bigg{)}+ 
\cos^2\alpha\; d\theta^2+d\alpha^2+\sin^2\alpha \; 
d\Omega^2_3\bigg{]} .
\eea
The boundary of this orbifold,where the dual gauge theory resides, is
actually $S^1/\Gamma\times S^3$. As we see because of the orbifold fixed 
points on the boundary, unlike the usual $AdS$ case, it is not possible
to extend over the circle. In other words, in this case we have closed 
time-like curves.

Similarly one can also consider orbifolds of $S^5$ \cite{ks, vafa}. 
In this case, however, 
there is only one choice for the half supersymmetric orbifold where the 
quotient space has a fixed circle with the $SU(2)\times U(1)$ symmetry 
remaining (out of the global $SO(6)$ rotation symmetry).
Therefore we have three kinds of  half supersymmetric orbifolds of 
$AdS_5\times S^5$. A similar classification can also be used for 
the (half SUSY) $AdS_{4,7}\times S^{7,4}$ orbifolds.

\section{Field theories duals to  $AdS_5\times S^5$ orbifolds}  

In this section we will discuss the field theories which are dual to the 
$AdS_5$ orbifolds discussed in the previous section. We will also compare the
results to the case of $S^5$ orbifold treated extensively in the literature 
\cite{ks,vafa}. For simplicity, we only consider the case of $A_n$ 
orbifolds i.e. $\Gamma=\bf{Z_n}$. 

For $\bf{Z_k}$ orbifolds of $S^5$ which leave a fixed $S^1$, one starts in 
the 
covering space with a gauge group $U(k N)$ and then projects out the
$\bf{Z_k}$ invariant components. If the $S^5$ is described by three
complex scalar fields $\phi_i, i=1,2,3$, then the projection is ensured by the 
condition:
\bea\label{co1}
\Omega~Z(x)~\Omega^{-1} = Z(x), 
\eea
\bea
\label{co2}
\Omega~\phi_j(x)~\Omega^{-1} = \omega \phi_j(x)
\eea
where we can choose $Z(x) = \phi_3(x)$ to denote the fixed  $S^1$, 
$\phi_j(x)$ the other two complex coordinates and
$\Omega = \mbox{diag}(1,\omega,~\cdots~,\omega^{k-1})$. The projection 
$\Gamma$ is global (is the same for any point $x$).
There is also an action on the gauge fields $A_{\mu}$ which implies 
that they can also be diagonalized in $k$ $N \times N$ blocks. 
The resulting spectrum is that of a four dimensional ${\cal{N}} = 2$ 
quiver theory 
with the gauge group $(U(N))^k$ and bifundamentals. The fields 
$A_{\mu},~Z$ together with the spinors form $k,~{\cal{N}} = 2$ vector 
multiplets and the bifundamentals form hypermultiplets. The beta function 
vanishes for each of the gauge groups $U(N)$ inside  $(U(N))^k$. 
In the supergravity side this corresponds to the existence of a fixed line 
corresponding to the dilaton for solutions of type $AdS_5 \times S^5/Z_k$ 
as stated in \cite{ks}. There are also twisted sectors which live on 
the fixed locus of the $Z_k$ action i.e. $AdS_5 \times S^1$, and their 
spectrum has been compared to the spectrum of chiral operators in field 
theory \cite{gukov}.

Now let us consider  the field theory dual to  
the $AdS_5$ orbifolds discussed in the previous section. 
To do that, we will review arguments of \cite{horov}. 
Their results, however, is more general than $S^3/\Gamma$ orbifold and is 
applicable to  gauge theories on 
$M/\Gamma$ manifolds where $M$ is any compact Riemann manifold and 
$\Gamma$ is any freely acting discrete isometry group. If we restrict 
ourselves to the case $\Gamma~=~\bf{Z_k}$, a field $\phi$ on
$M/\bf{Z_k}$ transforming under the adjoint action of a gauge group 
$U(kN)$ can be characterized as a pullback $s^{*} \tilde{\phi}$, 
by a local section $s~:~M/\bf{Z_k} \rightarrow M$, of a map 
$\tilde{\phi}~:~M \rightarrow C^{k~N} \otimes C^{k~N}$ transforming as 
\bea\label{con1}
\tilde{\phi} (u \cdot \gamma) = {\cal{R}}(\gamma^{-1}) \tilde{\phi}(u) 
{\cal{R}}(\gamma)
\eea
where $u$ belongs to $M$ and $\gamma$ to $\bf{Z_k}$ and 
${\cal{R}}(\gamma)$ is a representation of  $\bf{Z_k}$ and an element of 
$U(k)\times {\bf 1}_{N\times N}$. 

We can now state the difference between the $S^5$ and 
the $AdS_5$ orbifolds. Note that while the conditions 
(\ref{co1}),  (\ref{co2})
constrain the entries of the matrices $Z(x)$ (the RHS and LHS being at 
the same point), the condition (\ref{con1}) relate fields at different 
points $u$ and $u \cdot\gamma$. 
In other words, the orbifolding is defined by the ``twisted boundary 
condition'' for the $\phi$ field on $M/\bf{Z_k}$ as given in (\ref{con1}).
One may remove ${\cal{R}}(\gamma)$ elements by performing a gauge 
transformation. However, this gauge transformation is not globally 
defined, i.e. due to the non-trivial holomony of the gauge field $A$ the 
gauge transformation is not single valued when we go from one patch to 
another. Therefore we are forced to introduce a non-trivial flat 
connection \cite{horov}.
In short, the fields $\phi$ on $M/\bf{Z_k}$ in the presence of \flat
are the equivariant (as defined in (\ref{con1})) subclass of all the field 
defined on $M$.
% Besides the condition (\ref{con1}), there are some extra nontrivial 
%gauge transformations which relate $s^{*}_{\alpha} \tilde{\phi}$ to 
%$s^{*}_{\beta} \tilde{\phi}$ where $\alpha$ and $\beta$ refer to 
%different horizontal local sections $s_{\alpha}~:~U_{\alpha} \rightarrow 
%M$. Then the orbifold has different actions on the fields $\phi$ as 
%compared to the one on the fields $\tilde{\phi}$: it has a ``global'' 
%action on the fields $\tilde{\phi}$ which does not depend on the patches 
%covering $M/\bf{Z_k}$ and a ``local'' action on the fields $\phi$. 
This is to be compared with the case of the $S^5$ orbifold, where the 
constraints on the fields appear only when we go to the covering space and
those constraints are on the gauge group indices, not
on the $x$ coordinate.

In order to describe the field theory dual to $AdS_5$ orbifolds, 
we start with a theory on $R \times S^3$ i.e ${\cal N} = 4,~U(kN)$ 
theory with 3 scalar fields in the adjoint representation $\Phi_i, 
i=1,2,3$ together with a gauge field $A_{\mu}$. 
Let us focus on  the modifications after modding out by 
$\bf{Z_k}$, necessary to obtain a theory on $R \times S^3/{\bf Z_k}$. 
In the supergravity side, the number of modes per unit volume does 
not change and the holographic principle tells us that the number of 
modes per unit volume in the field theory should be the same too. 
In fact in \cite{horov} it is demonstrated that there exists a flat 
connection 
$A$ such that for every value of the Laplacian on $S^3$ (with zero 
connection) there exists solutions to $D^2_A \psi = \lambda \psi$ on 
$S^3/{\bf Z_k}$ and the number of such modes is
$1/k$ times the number of corresponding modes in  $S^3$. 
Since the volume of $S^3$ is also reduced by $1/k$ after the 
projection, the number of modes per unit volume remains the same. 

The action of ${\bf Z_k}$ breaks the Lorentz group in the field 
theory side and the theory is not conformal anymore. This is not an 
obvious statement and we now explain it. The conformality breaking can 
be traced back to either a non-vanishing beta function or to some
scale introduced in the theory. We use the latter to explain 
non-conformality of our theory. The argument is similar to that of 
\cite{hashimoto} where $U(N)$ field theory was obtained on one brane 
wrapped $N$ times on a circle of radius $L$. For the wrapped brane there 
are only $N$ types of open strings on a circle of length $N L$, which 
has lower energy states and corresponds to adding the flat connection in 
the gauge theory. The flat connection is proportional to $1/L$ so 
introduces a scale in field theory. The same appears in our case, 
for the $U(N)$ field theory on $S^3/{\bf Z_k}$ where the flat 
connection will depend on $1/\mbox{vol}( S^3)$ which becomes a scale 
in the field theory and this makes the field theory non-conformal. 
Another explanation for the non-conformality would be the non-zero 
value of the beta function. This is implied by the fact that when 
calculating the beta function, one has to use 
fields belonging to different patches, which introduce summation 
over extra phases in the beta function and those summations do not add up 
to zero (it is proportional to $\sin^2(2\pi/k)$.)

\section{Penrose limits and dual field theories}

Now we are ready to zoom-in onto a light-like geodesic and take the 
Penrose 
limit. In general we can choose this geodesic inside $AdS_5$ or inside 
$S^5$. For the geodesics inside $AdS_5$ we can either follow the 
radial direction $\rho$ or those which are parallel to the boundary 
(i.e. we boost along a circle inside $S^3$ part of $AdS_5$ space). 
It is not hard to see that these  paths inside the $AdS_5$ are 
``almost'' light-like and they  only become exact geodesics when we are
at strict $\rho=\infty$ which is the boundary. One can show that the 
Penrose machinery does not work for such geodesics which are not
formally in the space-time. More explicitly, if we take the $R\to\infty$ 
limit in this case and scale the other $AdS_5$ coordinates properly to 
keep 
metric components finite we end up with a metric which is not a solution 
of supergravity equations. Hence we only focus on the geodesics which are
in the $S^5$ part.

For the three possible half supersymmetric orbifolds mentioned above, 
one can take Penrose limits in four different ways: two  
corresponding to the $AdS_5$ orbifolds and two corresponding to $S^5$ 
orbifolds.
\footnote{We would like to discuss a possible subtlety which may arise in 
applying the Penrose method \cite{Penrose} to orbifold space-times.
By construction Penrose method is a procedure defined in classical general 
relativity (GR). However orbifold spaces, due to singularities, are not 
generally well-defined objects in classical GR. In the orbifold cases of 
our interest we focus on a geodesic which completely lies in the singular 
locus of the space (e.g. $\rho=0,\ \alpha=0,\ \theta=t$ of the metric 
(\ref{caseI})). Then, the scaling are so that we do not touch the orbifold 
structure. In such cases orbifolding and taking the Penrose limit 
commute. We would like to thank M. Rangamani for a discussion on this 
point.}     
The two sphere orbifold cases have been discussed in 
\cite{mohsen,orb,kehagias}, 
where it is shown that, if we choose our geodesic along the fixed 
circle of the 
orbifold, we get a half supersymmetric plane wave (which can be 
thought of as the orbifold of the maximally supersymmetric plane wave).
If we boost along the circle where the orbifolding is acting, we will
get maximally supersymmetric orbifold (in this case the Penrose limit
washes out the orbifolding information).
As for the two other possibilities, coming from $AdS_5$ orbifolds, 
the case of space-time orbifold (we have called this case 
one) has been briefly discussed in \cite{kehagias} and shown that
Penrose limit of this case leads to again an orbifold of maximally 
supersymmetric plane wave. However, the Penrose limit of the case two
has not been explored yet.

\subsection{Field theory dual to the Penrose limit of the spatial $AdS_5$ 
orbifold}

We start by reviewing the Penrose limit of the space $AdS_5$ 
orbifold \cite{kehagias}.
Consider the metric (\ref{caseI}) and perform the following Penrose 
limit
\bea\label{limitI}
x^+&=&{1\over 2}(t+\theta),\;\;\;\;\; x^-={1\over 2}R^2
(t-\theta)\ , \cr
\rho&=&{r\over R}\ ,\ \ \ \  \alpha={x\over R}, \ \ \ R\to\infty   
\eea
and $x,r,x^+$ and $x^-$ fixed. Taking the limit we find
\be
ds^2=-4dx^+dx^- -(r^2+x^2)(dx^+)^2+dr^2+dx^2+x^2d\Omega^2_3+r^2d\Omega'^2_3\ ,
\label{rrrr}
\ee
where the $d\Omega'^2_3$ is the metric for $S^3/Z_k$. 

As we see from the previous formulas, after taking the Penrose limit, 
one cannot distinguish whether the orbifolding was originally in the 
$AdS$ or sphere parts. At the level of supergravity or string theory, 
this is expected as we boost along the same direction in both cases 
for which we have an $S^3$ orbifold. This may be understood as follows: 
after taking the Penrose limit on $AdS_5\times S^5$ we find a solution with 
$SO(4)\times SO(4)$ symmetry and the two $SO(4)$'s are at the same 
footing, i.e. there is a $Z_2$ symmetry which exchanges them \cite{Z2}. 
{}From the 
$AdS$ point of view one of the $SO(4)$'s is coming as a subgroup of 
$SO(4,2)$, the $AdS_5$ isometry group, and the 
other as a subgroup of $SO(6)$ symmetry of $S^5$. However, this difference 
is washed out by the Penrose 
limit. Now one may orbifold by a ${\bf Z_k}$ subgroup of either of the 
$SO(4)$'s.
However from the field theory point 
of view, this tells us that a subsector of a conformal theory (the 
one for $S^5$ orbifold) is identical to a subsector of a non-conformal theory 
(the one for $AdS_5$ orbifold). 
%For the case $AdS_5 \times S^5$, the
%above interchange appears at the level of exchanging the two $SO(4)$
%groups which appear in the Penrose limit from the isometry group
%$SO(4,2) \times SO(6)$. 
%But in the non-orbifold case we only work 
%with conformal theory so the anomalous dimensions in the Penrose limit 
%can be compared \cite{gursoy}.

Our task here is to work out the explicit form of BMN operators in 
terms of the gauge theory on $R\times S^3/{\bf Z_k}$. Similar operators 
have been constructed in terms of quiver gauge theory operators 
\cite{{mohsen, orb}}. First we need to recognize what are the 
correspondents 
of the string light-cone momentum and energy. Although the full conformal 
group for the gauge theory on $R\times S^3/{\bf Z_k}$ is broken, we still 
have $U(1)\times (SU(2)\times U(1))$ subgroup of $SO(4,2)$, where the 
eigenvalues for the first $U(1)$ factor are the conformal dimension of 
the operators, $\Delta$. Besides that, we also have the full $SO(6)$ 
R-symmetry group. Following BMN, let us pick states carrying charge $J$ 
under a $U(1)$ factor of the R-symmetry. The Penrose limit
(\ref{limitI}) in the gauge theory side corresponds to $\Delta, J\to 
\infty$ while keeping
\bea
2p^-&=&i\partial_{x^+}=i(\partial_t+\partial_{\theta})=\Delta-J\cr &&\cr
2p^+&=&i\partial_{x^-}={i(\partial_t-\partial_{\theta})\over R^2}
={\Delta+J\over R^2}\;,
\nonumber 
\eea
fixed.

As the first step, we should identify the operators corresponding to vacuum for 
strings on plane wave orbifolds. In particular we note that we should have 
$k-1$ twisted and one untwisted vacuum states. 
To start with let us recall that gauge theories on $R\times S^3$ (after a 
Wick rotation) are related to theories on $R^4$ (or more precisely 
$R^4-\{O\}$) through radial quantization, where translation along the 
original time direction corresponds to dilatation (scaling) for the theory 
on $R^4$ 
and $t=-\infty$ to the origin of $R^4$,  $O$. 
\footnote{Note that the $R^4$ metric $dr^2+r^2d\Omega^2_3$ is conformal to
${dr^2\over r^2}+d\Omega^2_3$.}
Then it is clear that gauge theory on $R\times S^3/{\bf Z_k}$ is obtained 
through the radial quantization of the same gauge theory on $R^4/{\bf 
Z_k}$. Next, we recall that the BMN operator corresponding to vacuum is
\[
|vac\rangle \longleftrightarrow {\rm Tr} Z^J(x)|0\rangle|_{x\to 0}
\]
where $x$ is a point on $R^4$ and $Z$ is the complex scalar field whose 
phase corresponds to the 
direction along which we have boosted in the Penrose limit.
For the orbifold case, however, taking the $x\to 0$ limit is not trivial 
and depending on which patch of $R^4$ we start with, we will find 
$k$ different answers, corresponding to the $k$ vacua we were looking for.
These $k$ vacua are then related by orbifold action defined earlier in 
section 3. Explicitly, using (\ref{con1}) these different vacua are
\bea 
\label{vac1}
|vac\rangle_q \longleftrightarrow \mbox{Tr} [S^q Z^J],~~S = \mbox{diag}(1, 
e^{2 \pi i/k},\cdots, e^{2 (k-1) \pi i/k})
\eea 
where now $Z$ has been defined only for one patch. 
It is useful to compare the above to the case of $S^5$ orbifolds. In that 
case, having twisted vacua stems from the fact that the gauge group is
$(U(N))^k$ and hence there are $k$ $Z(x)$ fields (in the adjoint 
representations of each gauge factor). In terms of $kN\times kN$ matrices,
it has exactly the same form as (\ref{vac1}), but the ``twisting phase''
comes from a different origin \cite{mohsen}. For the $AdS$ orbifold 
case, this comes 
from the fact that all the fields, including $Z$, on different patches
are related by a non-trivial gauge transformations, while for the $S^5$ 
orbifold, that is just manifestation of having a gauge theory with
$k$ copies of $U(N)$ factor with the certain quiver matter content.
\footnote{This ``similarity'' can also be understood in the context of the
``holographic'' model proposed in \cite{bn}, where the boundary of the 
plane wave is one dimensional as a matrix model being built out of 
the lowest lying KK modes of the  SYM on an $S^3$ compactification. The 
discussion of 
\cite{bn} can be used for $S^3/{\bf Z_k}$ as well. Then, in this point of 
view it 
is evident that in the $S^3/{\bf Z_k}$ orbifold case we should have $k$ 
different vacua.} 

Given the vacuum states one may proceed with insertions of 
the covariant derivative on $R^4/{\bf Z_k}$,  
$D^q_a,\ a=1,2,3,4$  and $q=1,\cdots , k$ 
and other four scalars $\phi_{a'}$ into the string of $Z$'s with proper 
phases. 
Since the situation would be quite similar to that of $S^5$ orbifolds
we do not repeat them here. (They can be found in Refs.\cite{mohsen,orb}.)  
As an example, let us discuss the first level operators.
The field $Z$ is in the adjoint representation of the 
gauge group so the same discussion applies for the vector field. Therefore the 
quantities $D_{a} Z = \partial_{a} Z + [A_{a},Z]$ 
(where $D_{a}$ is the covariant derivative for the gauge theory on 
$R^4$ and $A_a$ is the gauge field) will also be defined only for one 
patch and it will
give $k$ different answers for the $k$ vacua. Therefore we can 
define the quantities:
\bea
Tr [S^q~D_{a} Z], S = \mbox{diag}(1, 
e^{2 \pi i/k},\cdots, e^{2 (k-1) \pi i/k})
\eea 
as the twist invariant operators with $\Delta - J = 1$.

With this construction it is clear that the BMN sector of the dual gauge 
theories for $AdS_5$ and $S^5$ orbifold cases should be equivalent and 
both 
should describe the strings on the half supersymmetric plane wave 
orbifolds.

It is also important to note that since the plane wave background obtained 
 from taking the  Penrose limit of $AdS$ is the same as the
one obtained from the $S^5$ orbifold and, moreover, since the latter one 
is dual to a subsector of a conformal theory, we might conclude that 
the conformal symmetry is restored for the BMN subsector of the gauge 
theory on $R\times S^3/{\bf Z_k}$.
This situation is very similar to the supersymmetry enhancement in the 
Penrose limit procedure \cite{klebanov,mohsen,orb}. 

As a straightforward generalization of the above discussions one may 
consider the orbifolds 
$AdS_5/\Gamma \times S^5/\Gamma'$, where
$\Gamma, \Gamma'$ are discrete subgroups which give ${\cal N} = 1$ 
theories. After the Penrose limit, it is easy to check that 
$AdS_5/\Gamma \times S^5/\Gamma'$ and $AdS_5/\Gamma' \times S^5/\Gamma$ 
become identical.

\subsection{Penrose limit of the $AdS$ space-time orbifolds}

The  $AdS$ space-time orbifolds have been discussed in \cite{bl} 
where it has been claimed that they come from 
the near horizon limits of brane configurations with pp-waves. These 
brane configurations can be 
related to  brane configurations probing Taub-NUT space by a series of 
S and T -dualities. 
\footnote{One should note that, in general, taking the near horizon limit of 
a given brane configuration does not commute with $T$-dualities.}
 
The dual field theories contain closed timelike curves and we do not have 
a clear picture of field theory with such a property. 
Nevertheless, we can take the Penrose limit by boosting
along the $\theta$ direction in five-sphere and expand 
the metric (\ref{caseII}) about $\rho=0, \alpha=0$, i.e.
\bea\label{limitII}
x^+&=&{1\over 2}({t\over k}+\theta),\;\;\;\;\; x^-={1\over 
2}R^2({t\over k}-\theta)\ , \cr
\rho&=&{r\over R}\ ,\ \ \ \  \alpha={x\over R}, \ \ \ R\to\infty   
\eea
and $x,r,x^+$ and $x^-$ fixed. Taking the limit, after the redefinition of
$\phi$ as $\phi-x^+$, we obtain
\be
ds^2=-4dx^+dx^- 
-(r^2+x^2)(dx^+)^2+dr^2+dx^2+x^2d\Omega^2_3+r^2d\Omega'^2_3\ ,
\ee
which is exactly the maximally supersymmetric plane wave without any
orbifolding. Note also that for the subsector of the gauge theory
dual to this background the conformal symmetry has also been restored.

One should note that unlike the maximally supersymmetric Penrose limit of 
the sphere orbifold, which for large $k$ leads to a plane wave with a 
finite
$x^-$ compactification radius \cite{orb}, here the story is different and, 
even for large $k$, $x^-$ is non-compact.

As we do not know the field theory dual to the space-time orbifold, we cannot 
give a precise interpretation on how to extract the conformal invariant 
subsector out of the  field theory with closed timelike curves. The closed
time-like  curves appear because we identify points on the time axis. 
{}From the form of the Penrose limit (\ref{limitII}) we see that, in the 
limit, 
the ends of the time-like curve are sent to infinity so the 
closed time-like do not remain closed anymore after the limit.
It would be interesting to clarify this aspect. 

\section{Rotating strings} 

The field theory results of \cite{bmn} relating the energy and 
the angular momentum can be mapped in the string theory
by using the results of \cite{metsaev,blau}. An alternative method was 
suggested in \cite{gubser, FT} (see also \cite{russo,tse1,Wadia} for 
more involved examples). It consists of choosing a particular 
pointlike string classical 
solution and expanding the $AdS_5 \times S^5$ action to obtain the 1-loop 
correction to the energy.

In this section we shall study the example of classical rotating solution 
which describes a folded closed string stretched along radial
direction of $AdS$ and rotating along the orbifold part of the 
space and moving along the large circle of $S^5$. The classical 
energy is a function of two angular momenta $S$ and $J$ but the problem is 
simpler because there is no mixing between the two. The one loop 
corrections to 
the energy allows us to determine terms in the expansion of the anomalous 
dimensions in the field theory side. The quadratic approximation gives the 
relation \cite{bmn,metsaev}

\be
E-J = \sum_{n=-\infty}^{n=\infty}\sqrt{1 + \frac{\lambda n^2}{J^2}}~N_n,
\ee 
and the quartic approximation allows calculating the 
$O(\frac{1}{\sqrt{\lambda}})$ terms. 

\subsection{Classical solution}

\subsubsection{ Space orbifold}
Let us consider the case where the orbifold is defined only in the
space like directions and the background is given by (\ref{caseI}). 
We will look at a rotating string solution which is stretched along radial
direction and rotates along $\phi$ and $\theta$ with the angular velocity
$\omega$ and $\nu$ respectively. The solution is given by
\be
t=\kappa \tau,\ \;\;\;\phi=\omega\tau,\ \;\;\;\;\theta=\nu \tau,
\;\;\;\;\ \rho=\rho(\sigma)=\rho(\sigma+2\pi),\;\;\;
%\delta=\delta_0={\rm constant}\; 
\ \ \alpha=0\ ,
\ee
and all the other angular variables set to be constant.
The Nambu action
\be
I=-{1\over 2\pi \alpha'}\int d\sigma^2\sqrt{-\det(G_{\mu\nu}
\partial_{a}X^{\mu}\partial_b X^{\nu})},
\label{NAMBUS}
\ee
becomes
\be
I=-4{R^2\over 2\pi}\int _0^{\rho_0} d\rho\sqrt{
(\kappa^2-\nu^2)\cosh^2\rho -({\omega^2\over k^2}-\nu^2)\sinh^2\rho}\;,
\ee
where
\be
\coth^2\rho_0={\omega^2/k^2-\nu^2\over \kappa^2-\nu^2}\;.
\ee
One should also note that $\rho(\sigma)$ is subject to the constraint 
coming from fixing the conformal gauge
\be\label{const}
\left({d\rho\over d\sigma}\right)^2=
(\kappa^2-\nu^2)\cosh^2\rho -({\omega^2\over k^2}-\nu^2)\sinh^2\rho\; .
\ee
For this action, the energy, spin and R-charge 
\be
E=-{\partial I\over \partial \kappa},\;\;\;\;\;\;
S={\partial I\over \partial \omega},\;\;\;\;\;\;
J={\partial S\over \partial \nu},
\label{FORM}
\ee
are obtained to be
\bea
E&=&{4R^2\over 2\pi}\kappa\int_{0}^{\rho_0} 
\frac{\cosh^2\rho\;d\rho}{\sqrt{(\kappa^2-\nu^2)
\cosh^2\rho-({\omega^2\over k^2}-\nu^2)\sinh^2\rho}}\;,
\cr &&\cr
S&=&{4R^2\over 2\pi}{\omega\over k^2}\int_{0}^{\rho_0} 
\frac{\sinh^2\rho\;d\rho}{\sqrt{(\kappa^2-\nu^2)
\cosh^2\rho-({\omega^2\over k^2}-\nu^2)\sinh^2\rho}}\;,
\cr &&\cr
J&=&{4R^2\over 2\pi\alpha'}\nu\int_{0}^{\rho_0} 
\frac{d\rho}{\sqrt{(\kappa^2-\nu^2)\cosh^2\rho-
({\omega^2\over k^2}-\nu^2)\sinh^2\rho}}.
\label{ESJ5}
\eea

One observes that 
\be
E={\kappa\over \nu}J+{\kappa\over \omega}k^2S\;,
\label{ESJ}
\ee
and eq.(\ref{const}) and the periodicity also imply another condition on 
the
parameters
\be
2\pi=\int_0^{2\pi}d\sigma=4\int_0^{\rho_0}\frac{d\rho}{
\sqrt{(\kappa^2-\nu^2)\cosh^2\rho-({\omega^2\over k^2}-\nu^2)
\sinh^2\rho}}\;.
\ee

It is useful to define ${\hat S}=kS$ and $\Omega=\omega/ k$ and we recognize 
that in terms of these new variables the situation is the same as $AdS_5
\times S^5$. Therefore one can read the energy dependence on $J$ and $S$
form the ones of \cite{FT}. By setting 

\[
\coth^2\rho_0={(\Omega^2-\nu^2)\over
(\kappa^2-\nu^2)}\equiv 1+\eta,\;\;\;\eta >0\;,
\]
one finds:

{\bf A) Short string}: $\eta\gg 1$. In this case the energy dependence on
$S$ and $J$ is given by

\bea
E^2 & \approx & J^2+2R^2\;kS,\;\;\;\;\;\;\;\;\;\;\;\;\;
{\rm for}\;\nu\ll 1\;,\cr &&\cr
E & \approx & J+kS+\frac{R^4}{2}\;{kS\over J^2},\;\;\;\;\;\;
{\rm for}\;\nu\gg 1\,.
\eea

{\bf B) long string}: $\eta\ll 1$. In this case one get
\bea
E & \approx & S+{R^2\over \pi}\ln{kS\over R^2}+\frac{\pi J^2}
{2R^2\ln{kS\over R^2}},\;\;\;\;\;\;\;{\rm for}\;
\nu\ll \ln{1\over \eta}\;,
\cr &&\cr
E & \approx & S+J+{R^4\over 2\pi^2J}\ln{kS\over J},\;\;\;\;\;\;\;\;
\;\;\;\;\;\;\;\;{\rm for}\;\nu\gg \ln{1\over \eta}\;.
\eea

\subsubsection{Space-time orbifold}

The background for space-time orbifold is given by (\ref{caseII}). 
Although there is a potential problem with the existence of closed 
time-like curves, one can formally  repeat the previous 
semi-classical analysis of the strings in the space-time orbifold as well, 
without entering to any problem with the singular fixed points. 
We 
shall look at a rotating string solution which is stretched along radial
direction and rotates along $\gamma$ and $\theta$ with the angular velocity
$\omega$ and $\nu$, respectively and is fixed at $\sigma={\pi\over 2}$.
The solution is given by
\be
t=\kappa \tau,\;\;\;\;\;\phi=\omega\tau,\;\;\;\;\;\theta=\nu \tau,
\;\;\;\;\;\rho=\rho(\sigma)=\rho(\sigma+2\pi)\;.
\ee
For this solution the Nambu action (\ref{NAMBUS}) reads
\be
I=-4{R^2\over 2\pi}\int_0^{\rho_0} d\rho\sqrt{({\kappa^2\over k^2}-\nu^2)
\cosh^2\rho -(\omega^2-\nu^2)\sinh^2\rho}\;,
\ee
where
\be
\coth^2\rho_0={\omega^2-\nu^2\over \kappa^2/k^2-\nu^2}\;.
\ee
Using this action one can find the energy, spin and 
R-charge as following
\bea
E&=&{4R^2\over 2\pi}{\kappa\over k^2}\int_{0}^{\rho_0} 
\frac{\cosh^2\rho\;d\rho}{\sqrt{({\kappa^2\over k^2}-\nu^2)
\cosh^2\rho-(\omega^2-\nu^2)\sinh^2\rho}}\;,
\cr &&\cr
S&=&{4R^2\over 2\pi}\omega\int_{0}^{\rho_0} 
\frac{\sinh^2\rho\;d\rho}{\sqrt{({\kappa^2\over k^2}-\nu^2)
\cosh^2\rho-(\omega^2-\nu^2)\sinh^2\rho}}\;,
\cr &&\cr
J&=&{4R^2\over 2\pi\alpha'}\nu\int_{0}^{\rho_0} 
\frac{d\rho}{\sqrt{({\kappa^2\over k^2}-\nu^2)\cosh^2\rho-
(\omega^2-\nu^2)\sinh^2\rho}}.
\eea
The periodicity also implies 
\be
2\pi=\int_0^{2\pi}d\sigma=4\int_0^{\rho_0}\frac{d\rho}{
\sqrt{({\kappa^2\over k^2}-\nu^2)\cosh^2\rho-(\omega^2-\nu^2)
\sinh^2\rho}}\;.
\ee
We define ${\hat E}=kE$ and ${\hat \kappa}=\kappa/ k$ and we recognize 
that in terms of these new variables the situation is the same as for $AdS_5
\times S^5$. Therefore, by using the results of \cite{FT}, the energy
dependence on 
$J$ and $S$ is

{\bf A) Short string}: $\eta\gg 1$. In this case the energy dependence on
$S$ and $J$ is given by
\bea
k^2E^2 & \approx & J^2 +2R^2\;S,\;\;\;\;\;\;\;\;\;\;\;\;\;
{\rm for}\;\nu\ll 1\;,\cr &&\cr
kE & \approx & J+S+\frac{R^4}{2}\;{S\over J^2},\;\;\;\;\;\;
{\rm for}\;\nu\gg 1\,.
\eea

{\bf B) Long string}: $\eta\ll 1$. In this case one get
\bea
kE & \approx & S+{R^2\over \pi}\ln{S\over R^2}+\frac{\pi J^2}
{2R^2\ln{S\over R^2}},\;\;\;\;\;\;\;{\rm for}\;
\nu\ll \ln{1\over \eta}\;,
\cr &&\cr
kE & \approx & S+J+{R^4\over 2\pi^2J}\ln{S\over J},\;\;\;\;\;\;\;\;
\;\;\;\;\;\;\;\;\;{\rm for}\;\nu\gg \ln{1\over \eta}\;.
\eea

\subsection{One-loop correction}

By knowing  the classical expressions for the energy, one can now 
proceed to compute the 1-loop correction to the classical energy 
spectrum for the above classical solutions. 
We will consider only the case of rotation in $S^5$ i.e. 
solutions with $\omega=0$, $\kappa=k\nu$ and $\rho=0$. Then we will expand 
the action  up to quadratic order for the following fluctuations: 
\be
t=k\nu\tau+{{\tilde t}\over R},\;\;\;\;\theta=\nu\tau+{{\tilde \theta}\over R},
\;\;\;\;\rho={{\tilde \rho}\over R},\;\;\;\;\alpha={{\tilde \alpha}\over R}\;.
\label{Fluc}
\ee
This leads to the following bosonic action for the quadratic fluctuations
for the background (\ref{caseII})
\bea
I_2&=&-{1\over 4\pi}\int d\sigma^2\bigg{\{}-{1\over k^2}\partial_a{\tilde t}
\partial^a{\tilde t}+\partial_a{\tilde \theta}\partial^a{\tilde \theta}+
\nu^2({\tilde \rho}^2+{\tilde \alpha}^2)+\partial_a{\tilde \alpha}
\partial^a{\tilde \alpha}+\partial_a{\tilde \rho}\partial^a{\tilde \rho}
\cr &&\cr &&
+ {\tilde \rho}^2\bigg{[}\partial_a{\tilde \delta}
\partial^a{\tilde \delta}+\sin^2\delta\;
\partial_a{\tilde \gamma}\partial^a{\tilde \gamma}+
+\cos^2\delta\;(\partial_a{\tilde \phi}\partial^a{\tilde \phi}+
-2\nu\partial_{\tau}\phi-\nu^2)\bigg{]}\cr &&\cr &&
+{\tilde \alpha}^2\;\partial_a\Omega_3\partial^a\Omega_3
\bigg{\}}\;.
\eea
This is the same action that is found by expanding the string action
in the maximally plane wave background. To see this consider the following
change of variable
\be
{{\tilde t}\over k}=x^++x^-,\;\;\;\;\;{\tilde \theta}=x^+-x^-\;,
\ee
then in the quadratic approximation and in the light cone gauge
where we we have $x^+=2\nu\tau$ one finds
\bea
I_2&=&-{1\over 4\pi}\int d\sigma^2\bigg{[}-4\partial_ax^+
\partial^ax^--{1\over 4}({\tilde \rho}^2+{\tilde \alpha}^2)
\partial_ax^+ \partial^ax^+
+\partial_a{\tilde \alpha}
\partial^a{\tilde \alpha}+\partial_a{\tilde \rho}\partial^a{\tilde \rho}
\cr &&\cr &&
+{\tilde \rho}^2\;\partial_a{\hat \Omega}_3\partial^a
{\hat \Omega}_3
+{\tilde \alpha}^2\;\partial_a\Omega_3\partial^a\Omega_3
\bigg{]}\;,
\eea
where
\be
\partial_a{\hat \Omega}_3\partial^a
{\hat \Omega}_3=\partial_a{\tilde \delta}
\partial^a{\tilde \delta}+\sin^2\delta\;
\partial_a{\tilde \gamma}\partial^a{\tilde \gamma}+
+\cos^2\delta\;\partial_a{\tilde \phi}\partial^a{\tilde \phi}\;,
\ee
and we have redefined $\tilde \phi$ and ${\tilde \phi}-x^+$. As we see this is 
exactly the string theory on the maximally SUSY plane wave background, 
which is also due to the fact that the Penrose  limit of the background
gives the  maximally plane wave. 
Now we can use the results of string theory on the maximally SUSY 
plane wave 
background to write down the corrections to the energy spectrum up to 
order of ${\cal O}({1\over R^2})$, i.e.
\be
kE-J={1\over \nu}\sum_{n=-\infty}^{n=\infty}\sqrt{n^2+\nu^2}\;N_n+
{\cal O}({1\over R^2})\;,
\ee 
where $N_n$ is the occupation number for 8 sets of the bosonic
oscillators.

\subsection{Higher order correction}

So far we have seen that in the semi-classical approach cannot distinguish
between the $AdS$ and AdS space time orbifold in leading quadratic order 
and
it would be natural to ask whether the higher order corrections 
could see any difference between the  $AdS$ and $AdS$ space-time orbifold. 
Also, a natural question is about the possible differences between the 
$S^5$ and $AdS_5$ orbifolds. 

In this section we will study the 
next order corrections to the action, which are of order ${1\over R^2}$, by
keeping the terms which are proportional to ${1\over R^2}$ in the 
expansion of the previous section for small fluctuations
around the classical solution (\ref{Fluc}). Expanding up to order 
${1\over R^2}$, we get:
\bea
I_4&=&-{1\over 4\pi R^2}\int d\sigma^2\bigg{[}-\rho^2\;\sin^2\gamma\;
\partial_a{\tilde t}\partial^a{\tilde t}-{\tilde \alpha}^2
\partial_a{\tilde \theta}\partial^a{\tilde \theta}
+{{\tilde \rho}^4\over 3}(\nu^2+\partial_a{\hat \Omega}_3
\partial^a{\hat \Omega}_3)\cr &&\cr &&
\;\;\;\;\;\;\;\;\;\;\;\;\;\;\;\;\;\;\;\;\;\;\;\;-{{\tilde \alpha}^4\over 3}
(\nu^2+\partial_a \Omega_3
\partial^a \Omega_3)\bigg{]}\;.
\eea
As compared to the $AdS$ case \cite{FT}, the action has an extra factor  
$\sin^2\gamma$ in front of $\partial_a{\tilde t}\partial^a{\tilde t}$.

Let us now consider the semi-classical analysis for the $AdS$ space 
orbifold and
compare the results to the ones for the $S^5$ orbifolds.
The classical solution is now given by a state with
high momentum in $S^5$ part of the string background and localized around
$\rho=0$. We study the small fluctuation around this classical solution
as follows
\be
t=\nu\tau+{{\tilde t}\over R},\;\;\;\;\theta=\nu\tau+{{\tilde \theta}\over R},
\;\;\;\;\rho={{\tilde \rho}\over R},\;\;\;\;\alpha={{\tilde \alpha}\over R}\;.
\ee
For this fluctuations the bosonic part of the string action in the background
(\ref{caseI}) up to ${1\over R^2}$ is given by
\bea
I_2^{(AdS/\Gamma)}&=&-{1\over 4\pi}\int d\sigma^2\bigg{[}-\partial_a{\tilde t}
\partial^a{\tilde t}+\partial_a{\tilde \theta}\partial^a{\tilde \theta}+
\nu^2({\tilde \rho}^2+{\tilde \alpha}^2)+\partial_a{\tilde \alpha}
\partial^a{\tilde \alpha}+\partial_a{\tilde \rho}\partial^a{\tilde \rho}
\cr &&\cr &&
\;\;\;\;\;\;\;\;\;\;\;\;\;\;\;\;\;\;\;+
{\tilde \rho}^2\partial_a(\Omega_3/\Gamma)\partial^a(\Omega_3/\Gamma)
+{\tilde \alpha}^2\partial_a{\Omega'}_3\partial^a{\Omega'}_3\bigg{]}\;,
\cr &&\cr 
I_4^{(AdS/\Gamma)}&=&-{1\over 4\pi R^2}\int d\sigma^2\bigg{[}-
{\tilde \rho}^2\partial_a{\tilde t}
\partial^a{\tilde t}-{\tilde \alpha}^2\partial_a{\tilde \theta}
\partial^a{\tilde \theta}-{{\tilde \alpha}^4\over 3}
(\nu^2+\partial_a {\Omega'}_3
\partial^a {\Omega'}_3)\cr &&\cr &&
\;\;\;\;\;\;\;\;\;\;\;\;\;\;\;\;\;\;\;\;\;\;\;
+{{\tilde \rho}^4\over 3}\left(\nu^2+\partial_a(
\Omega_3/\Gamma)
\partial^a(\Omega_3/\Gamma)\right)
\bigg{]}\;,
\label{Adsorb}
\eea
where $\Omega'_3$ is the $S^3$ inside of $S^5$ and the 
orbifold acts on the
$AdS$ part. 

To compare the results with the $S^5$ orbifold, we also need the results for 
the latter. The loop corrections in this case is very similar to that of 
$AdS_5\times S^5$ studied in \cite{FT} (see also \cite{Wadia}) and the 
corresponding expanded bosonic action is
\bea
I_2^{(S^5/\Gamma)}&=&-{1\over 4\pi}\int d\sigma^2\bigg{[}-\partial_a{\tilde t}
\partial^a{\tilde t}+\partial_a{\tilde \theta}\partial^a{\tilde \theta}+
\nu^2({\tilde \rho}^2+{\tilde \alpha}^2)+\partial_a{\tilde \alpha}
\partial^a{\tilde \alpha}+\partial_a{\tilde \rho}\partial^a{\tilde \rho}
\cr &&\cr &&
\;\;\;\;\;\;\;\;\;\;\;\;\;\;\;\;\;\;\;+
{\tilde \rho}^2\partial_a\Omega_3\partial^a\Omega_3
+{\tilde \alpha}^2\partial_a({\Omega'}_3/\Gamma)
\partial^a({\Omega'}_3/\Gamma)\bigg{]}\;,
\cr &&\cr 
I_4^{(S^5/\Gamma)}&=&-{1\over 4\pi R^2}\int d\sigma^2\bigg{[}-
{\tilde \rho}^2\partial_a{\tilde t}
\partial^a{\tilde t}-{\tilde \alpha}^2\partial_a{\tilde \theta}
\partial^a{\tilde \theta}+{{\tilde \rho}^4\over 3}
(\nu^2+\partial_a \Omega_3
\partial^a \Omega_3)\cr &&\cr &&
\;\;\;\;\;\;\;\;\;\;\;\;\;\;\;\;\;\;\;\;\;\;\;
-{{\tilde \alpha}^4\over 3}\left(\nu^2+\partial_a(
{\Omega'}_3/\Gamma)
\partial^a({\Omega'}_3/\Gamma)\right)
\bigg{]}\;.
\label{Sorb}
\eea

As we see the leading quadratic action in both (\ref{Adsorb}) and 
(\ref{Sorb}) is identical. This is, of course, expected because the
Penrose limit of both of them lead to the same plane wave. On the other 
hand in the ${1\over R^2}$ order we observe a difference of sign. 
This is not a very striking feature of the solution, because even in the 
$AdS_5 \times S^5$ case studied in \cite{FT} the contribution to the
quartic action of the angular parts of $AdS_5$ and $S^5$ come with opposite 
signs (see equation (4.15) of \cite{FT}). 
But in our case the different sign feature is more dramatic, as it alters 
the map between the Penrose limit of $AdS_5$ orbifold and the Penrose limit 
of $S^5$ orbifold, which were identical up to the quadratic correction. 
This is because for the quartic correction, we start probing beyond the 
plane wave limit.

\section{Discussions}

In the present project, we have given arguments for the identification 
of conformal subsectors of $S^5$ and $AdS_5$ orbifolds. As the case of
$S^5$ orbifold has been extensively studies before,
we have concentrate on the $AdS_5$ orbifold and have used arguments  
based on \cite{horov} in order  to identify the set of states in the two
subsectors.

Constructing the corresponding BMN operators for both $AdS_5$ and $S^5$
orbifolds, we explicitly showed that the BMN operators for both cases are 
identical in form, though different in interpretations in the dual 
gauge theories. Then, we also studied the equivalence between the Penrose 
limits of $AdS_5\times S^5$ orbifolds through semi-classical string theory 
computations. All these analysis confirms the existence of the $Z_2$ 
exchange symmetry between the two $SO(4)$ factors of the plane wave 
isometry group, in accordance with results of \cite{Z2}.

As we work with  $AdS_5$ orbifolds, it should be interesting to expand 
supergravity fields 
in terms of harmonics on $AdS_5$ (as in \cite{kim} for $S^5$) and 
to project out the harmonics as in \cite{ot}. Besides, the orbifold will imply
the existence of twisted sectors living on $S^1 \times S^5$ whose description
is similar to \cite{gukov}.

Here we also briefly discussed the Penrose limit of half 
supersymmetric ``time-like'' orbifolds and showed that it leads to 
the maximal supersymmetric plane wave geometry. This in particular implies 
that a subsector of the theory which has closed time-like curves is 
equivalent to BMN sector of a conformal theory. Noting that 
the BMN sector is a non-trivial (interacting) subsector,  it will be 
very interesting to see whether this observation can be used to extract 
information about the dual field theory living on the boundary of 
space-time $AdS$ orbifold. We also studied semi-classical description of 
strings in the $AdS$ space-time orbifolds. As we showed it seems that in 
this case, unlike the case of usual flat space, we do not have (UV) 
problems. This may in part be related to the fact that the number of 
orbifold images of a given particle when the particle moves to the fixed 
point is finite. 

Finally we would like to point out that similar to the 
discussions of \cite{LMS} one can consider ``parabolic'' orbifolds of 
$AdS$ space \cite{simon}. These orbifolds can also be half supersymmetric.  
Unlike the case of \cite{LMS2}, in the $AdS$ case the ``shift'' can only 
be a rotational twist. It would be very interesting to study the Penrose
limit and BMN duals of the theories on $AdS_5/\Gamma^+$ spaces discussed 
in \cite{simon}.

\vspace*{1cm}

{\bf Acknowledgments}

We would like to thank Keshav Dasgupta, Petr Horava, Mukund Rangamani and
Joan Simon for very useful discussions. The 
work of M. M. Sh-J~is supported in part by NSF grant
PHY-9870115 and in part by funds from the Stanford Institute for
Theoretical Physics. R.T. was supported by the DOE grant 
DE-AC03-76SF00098, the NSF grant PHY-0098840 and by the 
the Berkeley Center for Theoretical Physics.

\end{document}